\documentclass[useAMS,usenatbib]{mn2e}
\usepackage{graphicx}
\usepackage{epsfig}
\usepackage{epstopdf}
\usepackage{amssymb}
\usepackage{mathrsfs}
\usepackage{array}
\usepackage{url}

\title[Modelling CH$_{3}$OH masers] {Modelling CH$_{3}$OH masers: Sobolev approximation and accelerated lambda iteration method}

\author[A.V. Nesterenok] {A.V. Nesterenok $^{1}$ \thanks{E-mail:alex-n10@yandex.ru} \\
  $^{1}$ Ioffe Institute, 26 Polytechnicheskaya St., 194021 Saint Petersburg, Russia}

\begin{document}

\date{Accepted . Received ; in original form }
\pagerange{\pageref{firstpage}--\pageref{lastpage}} \pubyear{2015}
\maketitle

\label{firstpage}

\begin{abstract}
A simple one-dimensional model of CH$_3$OH maser is considered. Two techniques are used for the calculation of molecule level populations: the accelerated lambda iteration (ALI) method and the large velocity gradient (LVG), or Sobolev, approximation. The LVG approximation gives accurate results provided that the characteristic dimensions of the medium are larger than 5--10 lengths of the resonance region. We presume that this condition can be satisfied only for the largest observed maser spot distributions. Factors controlling the pumping of class I and class II methanol masers are considered.

\end{abstract}


\section{Introduction}

Intense maser transitions of the CH$_3$OH molecule are observed towards high-mass star-forming regions. The high brightness temperature of the maser emission permits us to observe them with the very long baseline interferometry (VLBI) technique, achieving both very high angular and velocity resolution. 

Methanol masers have been empirically divided into two classes \citep{Bartla1987, Menten1991b}. The class I methanol masers are often found apart from strong radio continuum and infrared sources and are associated with the shocked molecular gas \citep{Cyganowski2009, Voronkov2014}. The strongest and most common class I methanol masers are $4_{-1} \to 3_0$~E at a frequency of 36.2~GHz and $7_0 \to 6_1$~A$^+$ at 44.1~GHz \citep{Voronkov2012}. The class II methanol masers reside in close proximity to individual young stellar objects (YSOs). The strongest and most common class II methanol masers are observed from the transitions $5_1 \to 6_0$~A$^+$ at 6.7~GHz and $2_0 \to 3_{-1}$~E at 12.2~GHz. In addition to the strong and common methanol maser transitions, there are a large number which are observed to exhibit maser emission in a smaller number of sources \citep{Ellingsen2011}. The class II methanol masers are tracers of the high-mass star-formation while the class I masers are found in regions of both high- and low-mass star formation \citep{Xu2008, Chen2011, Kalenskii2013}.

The strongest methanol maser sources often show emission in multiple transitions and are projected against ultra-compact (UC) H\,{\sevensize II} regions, the example being W3(OH) \citep{Sutton2001, Moscadelli2003}. However, interferometric observations of the class II methanol masers at frequencies of 6.7 and 12.2~GHz indicate that, in most cases, the maser emission is not associated with any observable radio continuum emission that can be identified as a signature of an UC H\,{\sevensize II} region \citep{Walsh1998, Minier2001}. Many of the methanol masers are associated with sources within infrared dark clouds, which are believed to mark regions where high-mass star formation is in its very early stages \citep{Ellingsen2006}. These results have been interpreted in terms of class II methanol masers tracing an evolutionary phase which largely precedes the formation of an UC H\,{\sevensize II} region. 

Early theoretical models of methanol masers employed the collisional pumping mechanism \citep{Strelnitskii1981}. Subsequently, it was established that class I methanol masers have the collisional pumping mechanism, but class II methanol masers need a radiation field as a pumping source \citep{Walmsley1988, Zeng1990, Zeng1992, Cragg1992}. The ground and the first torsionally excited state of the molecule were included in early numerical models. These models could not explain the observed high brightnesses of class II methanol maser sources. The model of strong class II methanol maser at 12.2~GHz was proposed by \citet{Sobolev1994}. In their model, methanol molecules are excited to the first and the second torsionally excited states by the radiation from hot dust. Inverted populations result from the radiative and collisional cascade back to the ground state in the relatively cold gas. Subsequently, the model was extended to explore physical conditions necessary for the pumping of the 6.7~GHz maser line and other maser transitions \citep{Sobolev1997, Sobolev1997b, Cragg2002, Cragg2005}. 

Various techniques have been developed to yield a self-consistent set of level populations and radiation fields \citep{Grinin1984, Rybicki1984, Hubeny2001}. The escape-probability approach to the solution of radiative transfer problems in spectral lines in a medium with large velocity gradients was formulated for the first time by \citet{Sobolev1957, Sobolev1960}. In most cases, only the LVG, or Sobolev, approximation has been applied to evaluate level populations of the methanol molecule. As originally formulated, the Sobolev approximation ignores the effects of continuous opacity on the line intensity. \citet{Hummer1985} generalized the LVG approximation to include these effects. We used their results in our calculations.

The current study is aimed at modelling of the pumping mechanism of methanol masers. For the first time, the LVG approximation with the full treatment of continuum effects and the ALI method are used in the calculations of methanol level populations and line intensities. The zone of validity of the LVG approximation is discussed.

\section{Description of the model}
\subsection{Geometrical and physical parameters of maser clouds}
The class II methanol masers at 6.7 and 12.2~GHz were intensively studied by the VLBI technique. \citet{Moscadelli2003} reported on the results of VLBA observations of the 12.2~GHz methanol masers towards the UC H\,{\sevensize II} region W3(OH). The maser emission is observed as a collection of maser emission centres -- 'maser spots'. The sizes of maser spots in W3(OH) vary in the range 1--7~au with a mean size of 3~au. The maser spot spectra are very well reproduced by single Gaussian profiles with characteristic FWHM line widths of about 0.1--0.3~km~s$^{-1}$. The maser spectral analysis suggests that the 12~GHz masers in W3(OH) are unsaturated \citep{Moscadelli2003}. The emission of unsaturated masers is significantly narrowed compared with the line profile of the emission coefficient. The line narrowing factor is about 3--4 for the strong unsaturated masers \citep{Watson2002}. Thus, the dispersion in turbulent velocities in masing gas clumps is expected to equal 0.5--1~km~s$^{-1}$ to provide observed maser line widths. 

The maser spots are often clustered in groups -- 'maser features', emitting in contiguous frequency channels. It is generally believed that one maser feature corresponds to a distinct masing cloud of gas -- kinematical interpretation \citep{Moscadelli2011}. We took this point of view in our calculations. In some models, maser spots are assumed to be correlations in the distributions of physical parameters within a maser formation region which is larger by orders of magnitude than the spot size \citep{Sobolev1998}. As was pointed out in \citet{Sobolev2012}, the kinematical interpretation of the maser spot velocities should not be considered as contradictory to existence of correlations within extended masing regions.

\citet{Moscadelli2011} studied the milliarcsecond structure of class II methanol masers at a frequency of 6.7~GHz at high velocity resolution in four high-mass star-forming regions. Most of the detected 6.7~GHz maser features present an ordered (linear, or arc-like) distribution of maser spots on the plane of the sky, together with a regular variation in the spot local standard rest (LSR) velocity with position. The sky-projected sizes of maser features range from 5 to 50~au. Typical values for the amplitude of the LSR velocity gradients (defined in terms of the derivative of the spot LSR velocity with position) were found to be 0.05--0.1~km~s$^{-1}$~au$^{-1}$. These data are considered as an estimate of the gas velocity gradient in maser clouds.

To date, there are few VLBI observations of class I methanol masers \citep{Voronkov2012}. \citet{Matsumoto2014} carried out the VLBI imaging of the 44.1~GHz class I methanol maser towards the high-mass star-forming region \textit{IRAS} 18151--1208. The sizes of maser components lie in the range 5--20~au and brightness temperatures reach $10^{10}$~K.

\citet{Minier2005} presented multiwavelength studies of five methanol maser sites which are not directly associated with an UC H\,{\sevensize II} region. Each radio-quiet maser site is associated with a high-mass and luminous molecular clump. The spectral energy distribution for the most of the sources has a cold component (40--50~K) and a hot component (100--250~K). The hot component most likely probes the central region of the envelope around the luminous protostellar object. In addition, colder gas clumps seen only at mm wavelengths are found near the methanol maser sites. These clumps might represent an even earlier phase of high-mass star formation or, alternatively, the cold clumps might be clusters of low-mass YSOs \citep{Minier2005}. These data can be used to estimate the characteristic dust temperatures in class I and class II methanol masers.

The methanol production in gas-phase chemistry reactions is not effective \citep{Hartquist1995}. The methanol is produced on dust grains through hydrogenation of CO molecules \citep{Garrod2006}. It is hypothesized that the methanol abundance is enriched in maser regions due to the evaporation of icy grain mantles or due to the sputtering processes in shock waves \citep{Hartquist1995, Flower2010b, Yusef-Zadeh2013}. The fractional abundance of methanol ice is deduced from observations of the ice mantle composition and equals about $10^{-5}$ in interstellar clouds \citep{Whittet2011}. This estimate is considered as an upper limit for the methanol molecule abundance in the gas phase. 

\subsection{Calculational model}
\begin{figure}
\includegraphics[width=84mm]{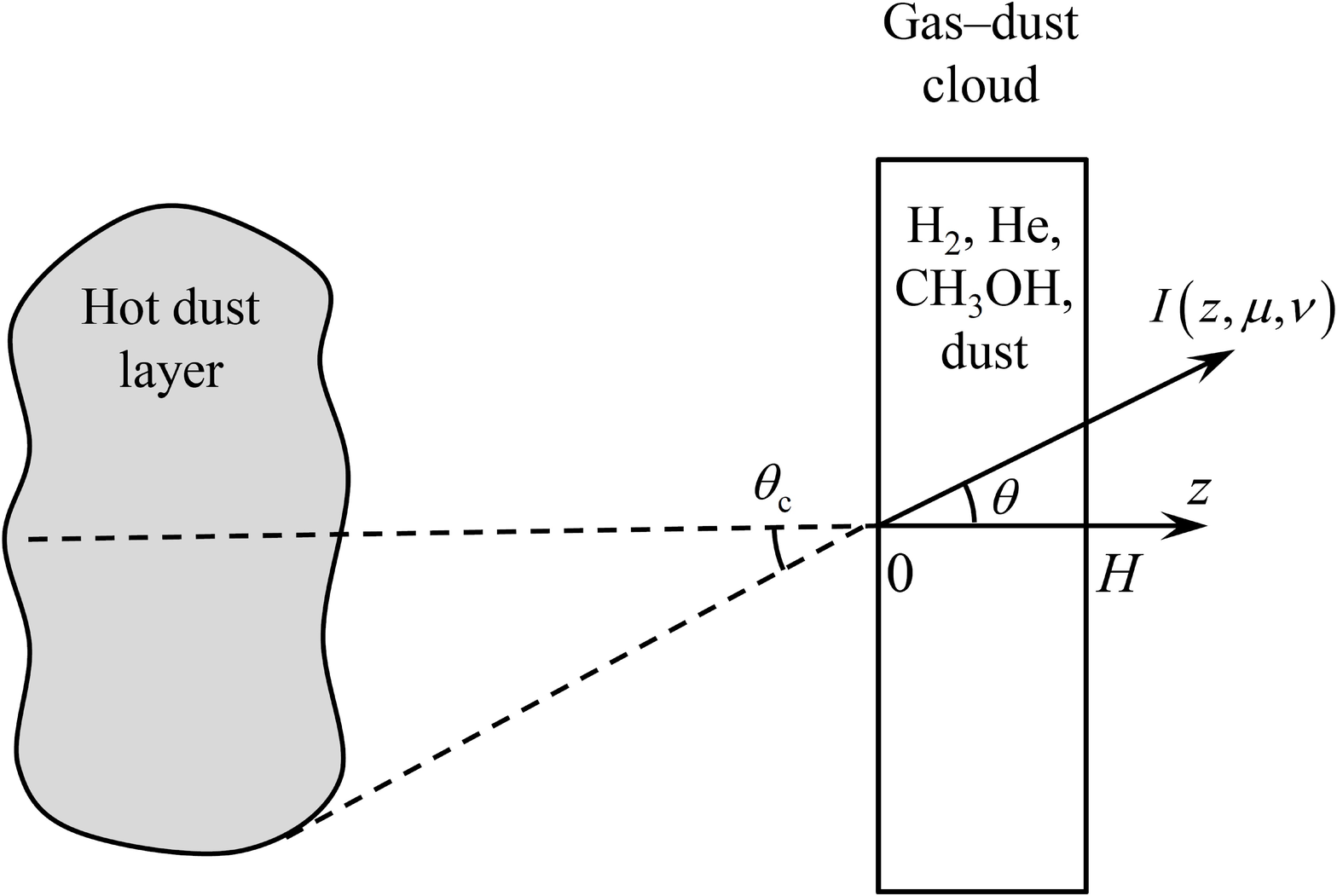}
\caption{Gas--dust cloud model.}
\label{fig1}
\end{figure}

We consider the one-dimensional model of a flat gas--dust cloud (see Fig. \ref{fig1}). The $z$ coordinate axis is perpendicular to the cloud plane. We assume that there is a constant gas velocity gradient along the $z$ axis, $k_{\rmn{v}} = dv/dz \geq 0$. The gas velocity $v(z) = 0$ at $z = 0$. Let $\mu$ to be the cosine of the angle between the $z$ axis and the radiation direction. The cloud consists of a mixture of H$_2$ and CH$_3$OH molecules, He atoms, and dust particles. The physical parameters of the cloud (the gas and dust temperatures, the number densities of atoms and molecules, and the dust abundance) were assumed to be independent of the coordinates.

In the calculations, we used the dust model from \citet{Weingartner2001} with the following parameters: the ratio of visual extinction to reddening $R_V = 3.1$, the carbon abundance in small grains $C/H = 55.8$~ppm, the dust--gas mass ratio 0.0081\footnote{http://www.astro.princeton.edu/\url{~}draine/dust/dustmix.html}. The dust emissivity was calculated in accordance with Kirchhoff's radiation law.

The model optionally includes the radiation of the external layer of hot dust at a temperature $T^{\rmn{ext}}_{\rmn{d}}$ with a filling factor $W_{\rmn{d}}$. Let $\mu_{\rmn{c}}$ to be the cosine of the angle characterizing the external radiation field (see Fig. \ref{fig1}). The radiation intensity of the external dust layer is: 

\begin{equation}
\begin{array}{l}
\displaystyle
I^{\rmn{ext}}_{\rmn{d}}(\nu, \mu) = B \left(\nu, T^{\rmn{ext}}_{\rmn{d}} \right) \lbrace 1 - \rmn{exp} \left[ -N^{\rmn{ext}}_{\rmn{H}} C_{\rmn{d}}(\nu) \right] \rbrace, \\ [5pt]
1 \geq \mu \geq \mu_{\rmn{c}},
\end{array}
\end{equation}

\noindent
where $B(\nu, T^{\rmn{ext}}_{\rmn{d}})$ is the Planck function, $C_{\rmn{d}}(\nu)$ is the dust opacity coefficient per H atom, $N^{\rmn{ext}}_{\rmn{H}}$ is the hydrogen column density of the external dust layer. The intensity $I^{\rmn{ext}}_{\rmn{d}}(\nu, \mu) = 0$ at $\mu < \mu_{\rmn{c}}$. In our model, the parameter $N^{\rmn{ext}}_{\rmn{H}}$ is independent of the radiation direction $\mu$ and equals $10^{23}$~cm$^{-2}$; the dust temperature $T^{\rmn{ext}}_{\rmn{d}} = 150$~K. The parameters of the external dust layer are similar to those, adopted by other workers \citep{Cragg2005}. The cosmic microwave background (CMB) radiation at a temperature of 2.7~K was taken into account in our model. The UC H\,{\sevensize II} background continuum radiation was not considered in our calculations. 

The model parameters are given in the Table \ref{table1}. Our list of physical parameters is based on the results of previous numerical models \citep{Cragg2005, McEven2014}. 

\begin{table}
\caption{Parameters of the model}
\begin{tabular}{p{4cm} >{\centering\arraybackslash}p{1.5cm} >{\centering\arraybackslash}p{1.5cm}}
\hline \\ [-1ex]
\qquad \qquad \qquad 1 & 2 & 3 \\ 
\hline \\ [-1ex]
Cloud thickness, $H$ & \multicolumn{2}{c}{30~au} \\ [5pt]
Number density of H$_2$, $N_{\rmn{H_2}}$ & \multicolumn{2}{c}{$5 \times 10^{6}$~cm$^{-3}$} \\ [5pt]
Number density of CH$_{3}$OH (A- and E-species), $N_{\rmn{m}}$ & \multicolumn{2}{c}{100~cm$^{-3}$} \\ [15pt]
Gas temperature, $T_{\rmn{g}}$ & 50--200~K & 150~K \\ [5pt]
Dust temperature, $T_{\rmn{d}}$ & 50~K & 150~K \\ [5pt]
Micro-turbulent speed, $v_{\rmn{turb}}$ & \multicolumn{2}{c}{0.5~km~s$^{-1}$} \\ [5pt]
Velocity gradient, $k_{\rmn{v}}$ & \multicolumn{2}{c}{0.05~km~s$^{-1}$~au$^{-1}$} \\ [5pt]
\hline \\
\end{tabular}

The physical parameters given in the columns 2 and 3 are favourable for the pumping of class I and class II methanol masers, respectively. 
\label{table1}
\end{table}

\section{Calculation of the CH$_3$OH level populations}
\subsection{Spectroscopic data and collisional rate coefficients}
Methanol molecules exist in two distinct nuclear-spin forms, related to the identity of the protons in the methyl (CH$_3$) group. In A-type methanol the spins of the three protons are 'parallel' and the resultant nuclear spin quantum number $I = 3/2$. The levels of A-type methanol are labelled by the '$+$' or '$-$' symbol related to the parity quantum number. In E-type methanol the spin of one of the protons is opposed to the spins of the other two, and $I = 1/2$. As E-type methanol exists in two degenerate forms, the statistical abundance ratio of symmetry species equals 1. Here, the A and E molecule species are assumed to be equally abundant.

Our calculations of methanol level populations took into account levels with angular momentum quantum number $J \leq 15$ and belonging to torsional states $v_{\rmn{t}} =$ 0, 1 and 2. The number of levels within each torsional state is equal to 256, making 768 levels in total for each of the symmetry states of the molecule. Energy levels and line strengths were taken from \citet{Mekhtiev1999}.

The rate coefficients for collisional transitions between methanol levels in collisions of methanol with He atoms and H$_2$ molecules were taken from \citet{Rabli2010, Rabli2010b, Rabli2011}. For the collisions of methanol with He atoms and para-H$_2$ molecules, \citet{Rabli2010, Rabli2010b} provided rate coefficients for rotational (torsionally elastic) transitions for the set of methanol levels and gas temperatures ($\leq 200$ K) considered in our calculations. For the ortho-H$_2$ as a collisional partner in CH$_3$OH--H$_2$ collisions, the rate coefficients were calculated by \citet{Rabli2010b} for transitions involving levels with $J \leq 9$ of the ground torsional state of the methanol molecule (100 levels). For $9 < J \leq 15$ and for transitions within torsionally excited states, the rate coefficients were assumed to be identical to those pertaining to collisions with para-H$_2$. This assumption will tend to underestimate the contribution of ortho-H$_2$ to the collisional transfer amongst these excited levels, as the ortho-H$_2$ rate coefficients tend to be larger than the para-H$_2$ \citep{Rabli2010b}. \citet{Rabli2011} computed rate coefficients for a restricted set of torsionally inelastic transitions in collisions of methanol with He atoms. For torsionally inelastic transitions induced by para-H$_2$, \citet{Rabli2011} suggested using the available rate coefficients for He atoms, and three times these values for collisions with ortho-H$_2$.

As the normal radiative and collisional processes do not change nuclear-spin of the molecule, the level populations of A- and E-type methanol were calculated independently. The effect of line overlaps on excitation of methanol masers was found to be unimportant \citep{Cragg2002}. \citet{Cragg2005} examined the effects of including more excited-state energy levels of methanol in the calculations of level populations (the torsionally excited states $v_{\rmn{t}} = 3, 4$ and CO-stretch vibrational mode). They found that these effects are significant only at high gas temperatures $T_{\rmn{g}} > 200$~K, at dust temperatures $T_{\rmn{d}} > 300$~K, and at large methanol column densities.

\subsection{Basic equations}
In the stationary case, the system of equations for the level populations $n_{\rmn{i}}$ is

{\setlength{\mathindent}{0pt}
\begin{equation}
\begin{array}{l}
\displaystyle
\sum_{k=1, \, k \ne i}^M \left[ R_{\rmn{ki}}(z)+C_{\rmn{ki}} \right] n_{\rmn{k}}(z) - \\ [15pt]
\displaystyle
- n_{\rmn{i}}(z) \sum_{k=1, \, k \ne i}^M \left[ R_{\rmn{ik}}(z)+C_{\rmn{ik}} \right]=0, \quad i=1,...,M-1, \\ [15pt]
\displaystyle
\sum_{i=1}^M n_{\rmn{i}}(z) = 1,
\end{array}
\label{stat_eqn}
\end{equation}}

\noindent
where $M$ is the total number of levels, $R_{\rmn{ik}}(z)$ is the rate coefficient for the transition from level $i$ to level $k$ through radiative processes, and $C_{\rmn{ik}}$ is the rate coefficient of collisional processes. The values of $R_{\rmn{ik}}(z)$ are given by

\begin{equation}
\begin{array}{l}
R_{\rmn{ik}}^{\downarrow}(z) = B_{\rmn{ik}}J_{\rmn{ik}}(z) + A_{\rmn{ik}}, \quad i > k, \\[10pt]
R_{\rmn{ki}}^{\uparrow}(z) = B_{\rmn{ki}} J_{\rmn{ik}}(z),
\end{array}
\nonumber
\end{equation}

\noindent
where $A_{\rmn{ik}}$ and $B_{\rmn{ik}}$, $B_{\rmn{ki}}$ are the Einstein coefficients for spontaneous emission and for stimulated emission and absorption, respectively; $J_{\rmn{ik}}(z)$ is the radiation intensity averaged over the direction and over the line profile. The radiative transfer in the plane-parallel geometry is considered. In this case, 

\begin{equation}
J_{\rmn{ik}}(z)=\frac{1}{2} \int\limits_{-\infty}^{\infty} d\nu \int\limits_{-1}^{1} d\mu \; \phi_{\rmn{ik}}(z,\mu,\nu) I(z,\mu,\nu)
\end{equation}

\noindent
where $I(z,\mu,\nu)$ is the intensity of radiation at a frequency $\nu$ in direction $\mu$ and $\phi_{\rmn{ik}}(z,\mu,\nu)$ is the normalized spectral line profile. The line intensity $I(z,\mu,\nu)$ is derived from the equation of radiative transfer \citep{Mihalas1984}:

\begin{equation}
\mu \frac{dI(z,\mu,\nu)}{dz} = -\kappa(z,\mu,\nu) I(z,\mu,\nu) + \varepsilon(z,\mu,\nu),
\label{rad_transf_eqn}
\end{equation}

\noindent
where $\kappa(z,\mu,\nu)$ and $\varepsilon(z,\mu,\nu)$ are the absorption and emission coefficients, respectively. The spectral profile of the emission and absorption coefficients in the laboratory frame of reference is

\begin{equation}
\displaystyle
\phi_{\rmn{ik}}(z, \mu, \nu) = \tilde{\phi}_{\rmn{ik}} \left[\nu - \mu \nu_{\rmn{ik}} v(z)/c \right]
\nonumber
\end{equation}

\noindent
where $\tilde{\phi}_{\rmn{ik}}(\nu)$ is the normalized spectral line profile in the co-moving frame of the gas, $\nu_{\rmn{ik}}$ is the transition frequency, $v(z)$ is the gas velocity along the $z$ axis. The line profile width is determined by the spread in thermal velocities of molecules and turbulent velocities in the gas--dust cloud:

\begin{equation}
\Delta\nu_{\rmn{ik}}=\nu_{\rmn{ik}}\frac{v_{\rmn{D}}}{c}, \quad v_{\rmn{D}}^{2} = v_{\rmn{th}}^2 + v_{\rmn{turb}}^2,
\nonumber
\end{equation}

\noindent
where $v_{\rmn{th}}$ is the most probable value of the thermal speed of the molecules, and $v_{\rmn{turb}}$ is the characteristic micro-turbulent speed in the cloud.

\subsection{Accelerated $\Lambda$--iteration method}
The detailed description of the ALI method for solving the system of master equations for molecule level populations coupled to the radiative transfer equation was given by \citet{Rybicki1991}. This method was employed in our modelling of the H$_2$O maser pumping in \citet{Nesterenok2013, Nesterenok2013b, Nesterenok2015, Nesterenok2014}. Here, the numerical scheme was modified in order to treat large velocity gradient of the gas in the cloud. The maximum value of the layer thickness $z_{\rmn{max}}$ is determined by the condition that the gas velocity change across this distance equals $0.1v_{\rmn{D}}$. The number of layers into which the cloud was broken down is $N \geq 100$ and depends on the value of the velocity gradient. The radiative transfer for the transitions with inverted level populations was treated in the way analogous to that used in \citet{Yates1997} -- the negative value of the line opacity was changed to positive value equal to $\zeta = 0.1$ of the absolute value of the opacity. The numerical scheme was tested using different values of the parameter $\zeta$.  

\subsection{Large velocity gradient approximation}
The equation of radiative transfer (\ref{rad_transf_eqn}) can be integrated for the plane-parallel medium with a monotonic velocity field \citep{Hummer1985}. In the Sobolev limit, 

\begin{equation}
z \gg \Delta z_{\rmn{D}}, \quad \Delta z_{\rmn{D}} = v_{\rmn{D}} \frac{dz}{dv},
\end{equation}

\noindent
where $z = 0$ corresponds to the cloud boundary. Characteristic dimensions of the medium are assumed to be much larger than $\Delta z_{\rmn{D}}$. Let us introduce parameters of a molecular line $i \to k$: 

\begin{equation}
\gamma = \frac{1}{\kappa_{\rmn{L}} \Delta z_{\rmn{D}}}, \quad \delta =  \frac{1}{\kappa_{\rmn{c}} \Delta z_{\rmn{D}}},
\end{equation}

\noindent
where $\kappa_{\rmn{c}}$ is the absorption coefficient of the dust at the line frequency, $\kappa_{\rmn{L}} \phi(x)$ is the line opacity coefficient, $\phi(x)= \rmn{exp}(-x^2)/\pi^{1/2}$, $x = (\nu - \nu_{\rmn{ik}})/\Delta\nu_{\rmn{ik}}$. The parameters $\delta$ and $\gamma$ are taken to be general, slowly varying, functions of $z$ coordinate. 

The mean intensity $J_{\rmn{ik}}(z)$ in the molecular line can be calculated \citep{Hummer1985}:

\begin{equation}
\begin{array}{l}
J_{\rmn{ik}}(z) = S_{\rmn{L}}(z) \left[1 - 2\mathscr{P}(\delta, \gamma) \right] + \\ [10 pt]
+ S_{\rmn{c}}(z) \left[ 1 - \mathscr{L}(\delta, \gamma, \tau_{\rmn{c1}}) - \mathscr{L}(\delta, \gamma, \tau_{\rmn{c2}})\right],
\end{array}
\label{line_int_lvg}
\end{equation}

\noindent
where $S_{\rmn{L}}$ is the line source function, $S_{\rmn{c}}$ is the source function in the continuum, $\tau_{\rmn{c1}}$ and $\tau_{\rmn{c2}}$ are the continuum optical depths to each of the cloud boundaries. The functions $\mathscr{P}(\delta, \gamma)$ and $\mathscr{L}(\delta, \gamma, \tau_{\rmn{c}})$ are one-sided loss probability functions for photons created by line and continuum processes, respectively:

\begin{equation}
\begin{array}{l}
\displaystyle
\mathscr{P}(\delta, \gamma) = \frac{1}{2} - \frac{1}{2} \int\limits_0^1 \frac{d\mu}{\gamma\mu^2} \int\limits_{-\infty}^{\infty} dx \phi(x) \times \\ [15 pt] 
\displaystyle
\times \int\limits_x^{\infty} dx'\phi(x') \rmn{exp} \left[ -\frac{1}{\gamma\mu^2} \int\limits_x^{x'} du \phi(u) - \frac{1}{\delta\mu^2}(x'-x) \right], \\ [15 pt]
\displaystyle
\mathscr{L}(\delta, \gamma, \tau_{\rmn{c}}) = \frac{1}{2} + \Gamma(\gamma, \tau_{\rmn{c}}) - \mathscr{P}(\delta, \gamma), \\ [15 pt]
\displaystyle
\Gamma(\gamma, \tau_{\rmn{c}}) = \frac{\gamma}{2}\int\limits_0^1 d\mu \mu^2 \rmn{e}^{-\tau_{\rmn{c}}/\mu} \left( 1 - \rmn{e}^{-1/\gamma\mu^2} \right).
\end{array}
\label{loss_prob_func}
\end{equation}

\noindent
When the continuous opacity is taken into account, the solution of the radiative transfer equation in the LVG approximation depends not only on the local parameters $\delta$ and $\gamma$, but on the cloud dimensions and the dust distribution in the cloud. Note that \citet{Hummer1985} did not consider an external radiation field.

The values of $\mathscr{P}(\delta, \gamma)$ were calculated for the grid of parameters $\gamma$ and $\delta$. The ranges for parameters $\gamma$ and $\delta$ were taken to be $[10^{-6},10^{14}]$ and $[10^{-2},10^{9}]$, respectively; the number of grid points per order of magnitude was taken to be 16. We used integration and interpolation algorithms published in the book by \citet{Press2007}. The relative accuracy of the integration was set to be $10^{-5}$, the bicubic interpolation within a grid square was used. 

The system of statistical equilibrium equations (\ref{stat_eqn}) with the line intensity given by (\ref{line_int_lvg}) was solved iteratively. The iterative series was treated in the same way as in the ALI method \citep{Nesterenok2015}. The convergence criterion for the iterative series was the condition on the maximum relative change in level populations for two successive iterations $< 10^{-4}$.

\begin{figure}
\includegraphics[width=84mm]{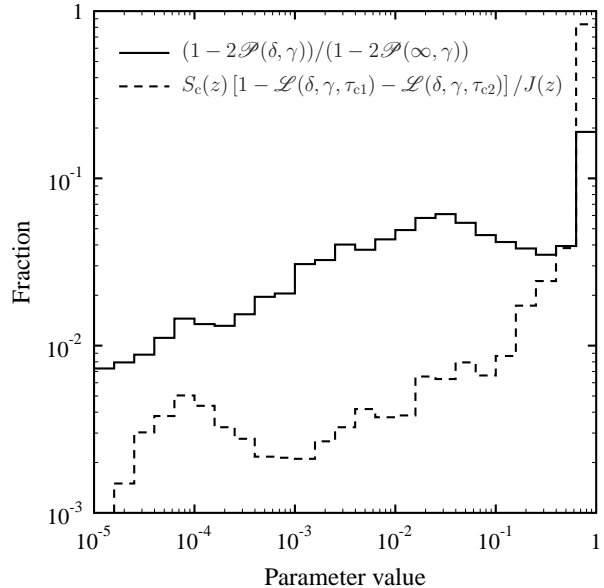}
\caption{The distribution of the parameters characterizing the dust effect on the line intensity. The number of bins per order of magnitude equals 5 on the x-axis. The y-axis shows the fraction of molecular lines that have the parameter value within the one bin. The results correspond to $z = H/2$.}
\label{fig2}
\end{figure}

\section{Results}
\subsection{LVG approximation and ALI method}
The physical parameters of the gas--dust cloud adopted in the calculations are those favourable for the pumping of the class II methanol masers (column 3 of the Table \ref{table1}). But the radiation from the external dust layer and the CMB radiation were not considered in the calculations (the LVG approximation equations used in our work do not account for external radiation fields). 

The effect of the continuous opacity on the line intensity in the LVG approximation was studied. The first term of the intensity expression (\ref{line_int_lvg}) was considered: the ratio of the term with the dust opacity taken into account to the term without taking into account the dust opacity was calculated for all radiative transitions of A- and E-species of methanol. The distribution of the parameter is shown in Fig. \ref{fig2} (solid line). Most of molecular lines are optically thin in the direction perpendicular to the cloud plane and have values $\gamma > 1$. The values of the one-sided loss probability function $\mathscr{P}(\delta, \gamma)$ are close to 0.5 in this case. However, the difference $1-2\mathscr{P}(\delta, \gamma)$ is very sensitive to the dust opacity. As a consequence, the first term in the intensity expression (\ref{line_int_lvg}) is reduced by more than an order of magnitude for a significant fraction of molecular lines when the dust opacity is taken into account. The ratio of the second term of the intensity expression (\ref{line_int_lvg}) to the total intensity is considered (dashed line in Fig. \ref{fig2}). It is seen that the parameter is close to unity for most of the molecular lines. The dust emission determines the radiation intensity for these molecular lines. 

\begin{figure}
\includegraphics[width=84mm]{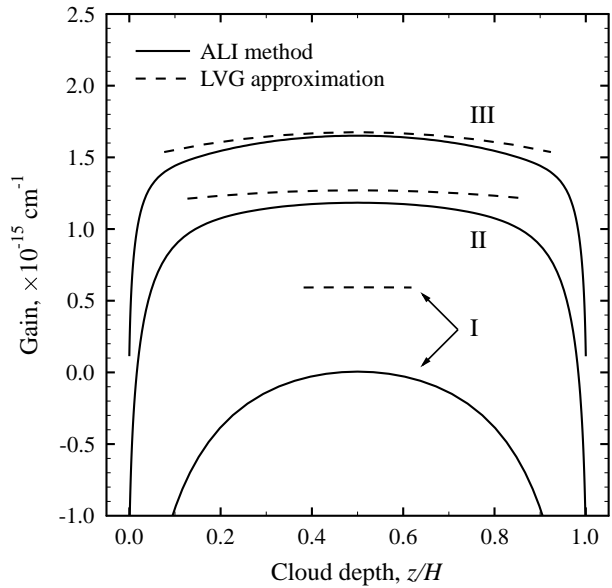}
\caption{The gain of the 6.7~GHz $5_1 \to 6_0$~A$^+$ maser line as a function of the cloud depth. The gain values are given at $\Delta z_{\rmn{D}} \leq z \leq H-\Delta z_{\rmn{D}}$ for the results of the LVG approximation.  The results are presented for three values of the cloud height: (I) 30~au; (II) 90~au; (III) 150~au. At the parameters in question, $\Delta z_{\rmn{D}}$ = 11.5 au.}
\label{fig3}
\end{figure}

The expression for the gain of the transition $i \to j$ at the line centre in a direction along the cloud plane is

\begin{equation}
\displaystyle
\gamma_{\rmn{ij}}(z)=\frac{\lambda^2 A_{\rmn{ik}} N}{8 \pi \sqrt{\pi} \Delta \nu_{\rmn{ij}}} \left[n_{\rmn{i}}(z)-\frac{g_{\rmn{i}}}{g_{\rmn{k}}}n_{\rmn{k}}(z) \right],
\nonumber
\end{equation}

\noindent
where $N$ is the number density of molecules (A- or E-symmetry species of methanol). Fig. \ref{fig3} shows the dependence of the gain of the maser line at 6.7~GHz on the cloud depth calculated by means of the ALI method and the LVG approximation. There is a significant discrepancy in the results of two methods at the cloud height $H = 30$~au. The parameter $H/\Delta z_{\rmn{D}}$ equals 2.6 in this case. The gain has negative values at almost all cloud depths according to accurate calculations, while the LVG approximation provides high positive values of the maser gain. There is an agreement between the results of two techniques at large cloud height -- the difference between the gain values at the cloud centre is about 10 per cent at $H = 90$~au ($H/\Delta z_{\rmn{D}}$ = 7.9) and about 2 per cent at $H = 150$~au ($H/\Delta z_{\rmn{D}}$ = 13). 

One can see in Fig. \ref{fig3} that the maser gain depends significantly on the cloud height. The larger the cloud height, the larger the dust optical depth and the radiation from dust is more intense. This makes the pumping of class II methanol masers more effective as these masers have a radiative pumping mechanism. At the model parameters in question ($H = 30$~au), the dust optical depth perpendicular to the cloud plane equals about 0.05 at wavelengths of 20--30~$\umu$m -- the approximate wavelengths of the pumping photons. Despite the dust optical depth being low, the effect of the dust intermixed with the gas on the maser pumping is significant. This effect is also discussed by \citet{Sutton2001}.

The simulation results for other maser transitions are analogous to those for the 6.7~GHz maser line shown in Fig. \ref{fig3}. The gain values calculated by means of the LVG approximation are higher than the results of accurate radiative transfer calculations for maser transitions which have radiative pumping mechanism (class II methanol masers). The gain increases with increasing cloud height for these maser lines. According to our calculations, the results of the LVG approximation are lower than accurate results for the collisionally pumped masers (class I methanol masers). For these masers, the gain decreases with increasing cloud height.

\subsection{Class I and class II methanol masers}
The results of the calculations of methanol level populations using the ALI method are presented in this section. The radiation from external dust layer (for class II methanol masers) and the CMB radiation were taken into account.

\begin{figure}
\includegraphics[width=84mm]{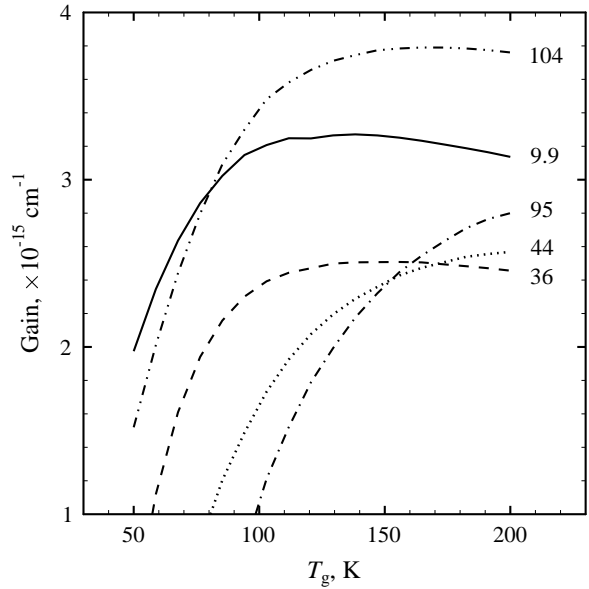}
\caption{The gain of the maser lines as a function of gas temperature. Only class I methanol masers are considered: $9_{-1} \to 8_{-2}$~E at a frequency of 9.9~GHz, $4_{-1} \to 3_0$~E at 36.2~GHz, $7_0 \to 6_1$~A$^+$ at 44.1~GHz, $8_0 \to 7_1$~A$^+$ at 95.2~GHz, and $11_{-1} \to 10_{-2}$~E at 104~GHz. The line frequency in GHz is indicated near each curve.}
\label{fig4}
\end{figure}

Fig. \ref{fig4} presents the dependence of the cloud-averaged gain of several class I methanol masers on the gas temperature. The physical parameters are given in the column 2 of the Table \ref{table1}. The gain of maser transitions increases with increasing gas temperature that is a signature of the collisional pumping mechanism. Our results are similar to the findings by \citet{McEven2014}. The gain does not depend on the abundance of cold dust in our model. 

\begin{figure}
\includegraphics[width=84mm]{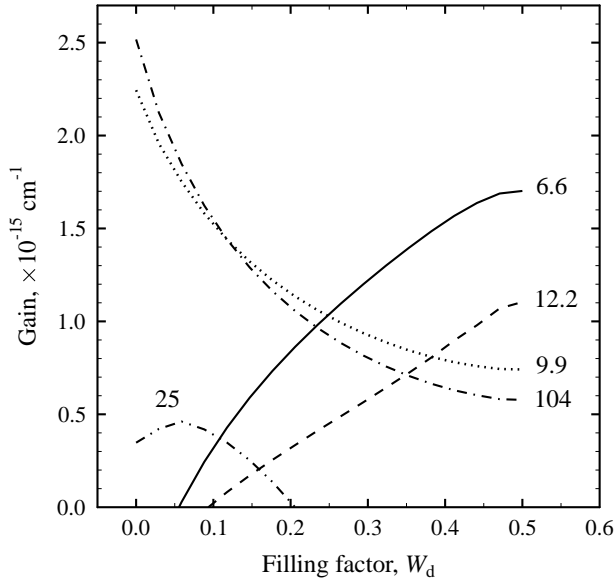}
\caption{The gain of the maser lines as a function of the filling factor of the external dust layer. The lines in question are: $5_1 \to 6_0$~A$^+$ at a frequency of 6.7~GHz, $2_0 \to 3_{-1}$~E at 12.2~GHz, $9_{-1} \to 8_{-2}$~E at 9.9~GHz, $5_{2} \to 5_{1}$~E at 25~GHz and $11_{-1} \to 10_{-2}$~E at 104~GHz.}
\label{fig5}
\end{figure}

Fig. \ref{fig5} shows the dependence of the cloud-averaged gain of several maser transitions on the filling factor of the external dust layer $W_{\rmn{d}}$. The physical parameters are presented in the column 3 of the Table \ref{table1}. The gain of class II methanol masers at frequencies of 6.7 and 12.2~GHz increases with increasing filling factor of the external dust layer. The opposite is true for the class I masers at frequencies of 9.9, 25 and 104~GHz. The collisional and radiative pumping mechanisms are competitive: the strong infrared radiation quenches collisionally pumped class I masers and strengthens class II masers. Nevertheless, masers of different types can coexist in the same maser region -- the gain of some class I methanol masers can be high in the strong infrared radiation field. Our results are consistent with the findings by \citet{Voronkov2005}.  

The length of the amplification region along the line of sight has to be $L \approx 20H = 600$~au to provide optical depth $\tau \gtrsim 20$ in strong maser lines in our model. According to our estimates, methanol masers become saturated at optical depths of about 15--20 (the beaming angle of the maser radiation was taken to equal $\Delta \Omega /4\pi = 10^{-3}$, the background radiation was not considered). The excitation temperatures of maser levels lie in the range 1--100~K. Thus, the model can account for high brightness temperatures $T_{\rmn{b}} \sim 10^9 - 10^{10}$ K of both class I and class II methanol masers, but large beaming factors $L/H \gtrsim 20$ are required. Note, the high gas-phase abundance of the methanol molecule $x_{\rmn{m}} = 10^{-5}$ was taken in our model to get high maser gain values.

\section{Discussion}
In the case of the large velocity gradient, the photointeraction at an arbitrary point of the medium is determined by the radiative coupling to its local neighbourhood -- the resonance region. The length of the resonance region $\Delta z_{\rmn{D}}$ depends on the half-width of the absorption coefficient profile and the velocity gradient at the given point. The LVG approximation gives exact solutions for the radiative transfer problem provided that the characteristic dimensions of the medium are much larger then the length of the resonance region \citep{Grinin1984}. The radiation intensity in molecular lines is determined by the loss probability functions that photons emitted at a point either escape from the resonance region or are absorbed in the continuum. The absorption by the continuum is an essential mechanism for photon loss in a dusty medium. Moreover, the dust emission can take a part in the pumping of the masers. We showed that the continuous opacity plays an important role in determining radiation intensity and must be taken into account accurately. Generally, the effects of the continuum emission and the absorption by the continuum within the resonance region were not considered or simplified assumptions were done in the solution of radiative transfer problem in high-speed flows \citep{Castor1970b, Deguchi1981}.

The maser emission arises from compact regions with the sizes of a few tens of au. We assume that each maser feature corresponds to a distinct gas--dust cloud with sizes comparable to the observed sizes of the emitting region. In this case, the dimension of the emitting region on the plane of the sky can be taken as a characteristic distance of changes of physical parameters of the medium such as gas density or molecule abundance. According to our calculations, the results of the LVG approximation and the ALI technique agree at large cloud height, $H > 5 \Delta z_{\rmn{D}}$. This condition can be satisfied only for the largest observed maser features, $H > 50$~au (at the velocity gradient and the micro-turbulent speed in question). In this case, the escape probability method gives accurate results in deep cloud layers but fails in outer layers. The results of the LVG approximation significantly differ from the results of accurate calculations at small cloud height $H \simeq 2\Delta z_{\rmn{D}}$ (see Fig. \ref{fig3}). In this case, the calculations based on the LVG approximation give wrong constraints on physical parameters in the maser source. For example, the LVG approximation calculations will reproduce the results shown in Fig. \ref{fig3} for the ALI method at small cloud height, if the gas density is taken two times lower than the value given in the Table \ref{table1}.

At small cloud height, the real loss probability functions are higher than the values given by equations (\ref{loss_prob_func}) \citep{Rybicki1984}. In this case, the radiation intensity in molecular lines is overestimated if equations (\ref{line_int_lvg},\ref{loss_prob_func}) are used. It affects gain values of the radiatively and collisionally pumped masers: the gain values are overestimated for radiatively pumped masers and underestimated for collisionally pumped masers.

\citet{Strelnitskii1981} and \citet{Sobolev1983} suggested a model of methanol masers observed towards the Orion molecular cloud. The collisional pumping mechanism was considered in their model in which the cold dust plays a key role in the pumping of the masers. The presence of cold dust would significantly enhance the level population inversion if the optical depth in the pumping lines is much greater than unity. In our model, most of ro-vibrational lines of the methanol molecule have low optical depth, and the probability for a photon to escape from the resonance region belongs to the interval $0.1-1$. As a result, the absorption of radiation by cold dust plays little role in the collisional pumping of class I methanol masers. 

Modelling of the maser pumping can provide estimates of the physical conditions in the maser regions \citep{Sobolev2012}. The presence and the absence of different maser transitions towards specific YSO can potentially be used to trace changes in physical conditions and the evolutionary state of the source \citep{Breen2010}. It is believed that the sources showing rare class II methanol maser transitions are the most evolved sources traced by class II methanol masers and arise just prior to the switch off of the methanol masers of this class \citep{Ellingsen2011}. A model attempting to simultaneously fit all of the observed methanol masers would have to take into account a source structure and possible saturation effects \citep{Sutton2001, Sobolev2002}. The LVG approximation is a local treatment of radiative transfer which does not take into account possible strong spatial variations of physical parameters in the source. The radiation transfer methods such as ALI method do not have this deficiency and the spatial variations of physical parameters can be accounted for in the calculations. The maser saturation has to be taken into account in modelling of strong maser sources with brightness temperatures $T_\rmn{b} \gtrsim 10^{10}$~K. The next step in the maser pumping simulations could be accurate radiative transfer calculations using a specific physical model of the source -- the time-dependent model of the photo-dissociation region or the interstellar shock wave model. 

\section{Conclusions}
A simple one-dimensional model of CH$_3$OH maser is considered. The following conclusions have been drawn:

(i) The LVG approximation calculations reproduce the results of accurate radiative transfer calculations at large cloud heights and high velocity gradients: the cloud height has to be of the order of or greater than 5-10 lengths of the resonance region. 

(ii) At small cloud height, the LVG approximation underestimates the photon loss probability functions. It can affect significantly gain values of the radiatively and collisionally pumped masers.

(iii) The absorption of radiation by dust and the dust emission play an important role in determining radiation intensity and must be taken into account accurately. 

(iv) The model presented can account for high brightness temperatures $T_{\rmn{b}} \sim 10^9 - 10^{10}$~K of the class I and class II methanol masers.

\section*{Acknowledgements}
This work was supported by the Russian Foundation for Basic Research (project no. 14-02-31302), the Program of the President of Russia for Support of Leading Scientific Schools (project no. NSh-294.2014.2). I am grateful to David Flower for providing the collisional rate coefficients for the CH$_3$OH--H$_2$ collisions. I thank Dmitry Kokorin for the help in C++ programming. I also thank the anonymous referee for valuable comments. The calculations were performed on the supercomputer of the Saint Petersburg branch of the Joint Supercomputer Center of the Russian Academy of Sciences\footnote{http://scc.ioffe.ru/}.

\bibliography{interstellar_medium_references}

\end{document}